\documentclass[12pt]{article}
\usepackage[usenames,dvips]{pstricks}
\usepackage{color}
\usepackage{amssymb}
\usepackage{footnote}
\usepackage{longtable}
\usepackage{verbatim}
\usepackage{array,multirow}
\usepackage{titlesec}
\usepackage{physics}
\usepackage{lineno}
\usepackage{url}
\usepackage{placeins}
\usepackage{float}

\usepackage{graphicx}
\usepackage{caption}
\include{symbols}

\def\baselinestretch{1.0}

\newrgbcolor{maroon}{.45 0.1 0.1}
\begin{document}
\pagestyle{plain}

\begin{center}
{\Large\bf Tungsten Thermionic Emission as a Gauge for Low Pressures of Cesium Vapor}
\end{center}

\vskip-0.5in

\begin{center}
Jo\~ao F. Shida, Fangjian Wu, Eric Spieglan\\
{\it Physics Department and Enrico Fermi Institute, the University of Chicago}\\
 Mesut \c{C}al{\i}\c{s}kan\\
{\it Physics Department and Kavli Institute for Cosmological Physics, the University of Chicago}\\
\today\\
\end{center}


\begin{abstract}
Heated metal filaments under electric fields and low pressures of alkali metal gas eject electrons by thermionic emission as a function of the pressure of the gas and the temperature of the filament. To explore this process in a program to develop large-area alkali metal photocathodes, we have designed and built a gauge following the studies of Taylor and Langmuir~\cite{sinclair, Taylor}. We present proof-of-principle measurements of the thermionic emission of a tungsten filament in cesium vapor. We describe a second generation design that corrects flaws in the first gauge.
\end{abstract}

%

\newpage
\section{Introduction}
\label{introduction}

Alkali metals adsorb onto metal surfaces, with a concentration and structure dependant on the temperature of the surface and the pressure of the metal vapor. The adsorption increases the thermionic emission of the surface compared to bare metal. We show that by measuring the thermionic emission as a function of the temperature of a tungsten filament, one can infer the pressure of cesium vapor around the filament.

\par 
According to Richardson`s law, the thermionic current is solely determined by the work function of the surface and its temperature:
\begin{equation}
    J=\lambda_R A_0 T^2 e^{-W/k_B T}
\end{equation}
The work function changes with the coverage of adsorbed cesium atoms. For coverages between 0 and 0.66, the adsorbed atoms lose an electron, creating an ionic charge layer which diminishes the work function and increases thermionic emission ~\cite{Layer}. 

For coverages above 0.66, the work function is low enough so that cesium ions that adsorb do not lose an electron, creating a neutral layer. To accommodate the larger radius of the neutral atoms, this layer is rotated 30 $^{\circ}$ about the normal ~\cite{Structure} relative to the ionized layer below. The work function approaches that of cesium metal, suppressing thermionic emission.


\par
\begin{figure}[H]
  \centering
  \includegraphics[height = 350pt]{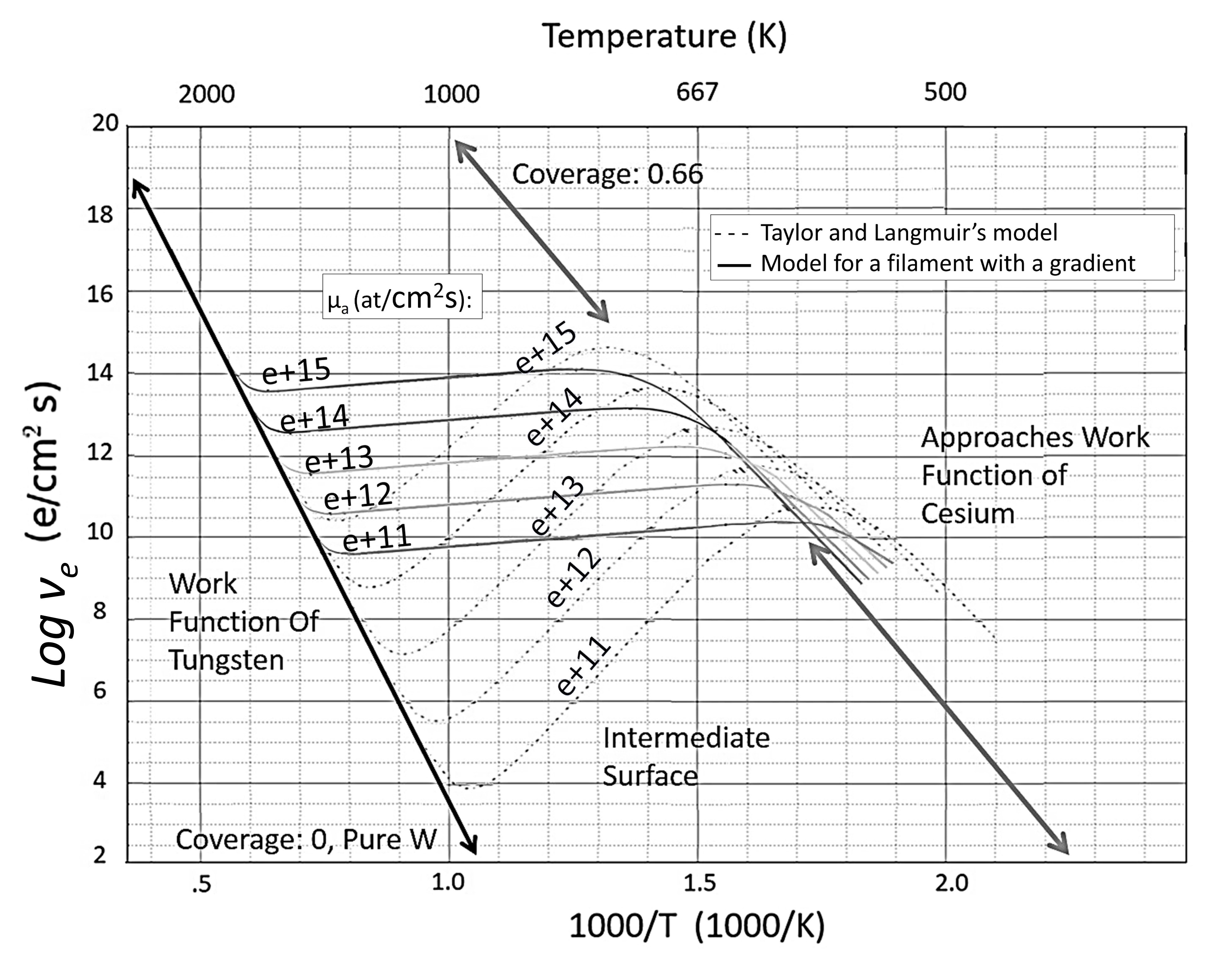}
  \caption{The electronic thermionic emission calculated with Taylor and Langmuir`s measurements is shown as dotted lines for 5 atom flux values of cesium proportional to vapor pressures (indicated on top of the curves, e.g. e+11). To account for the thermal gradient of the filament in our gauge, we have corrected the Taylor and Langmuir model by integrating over the filament, with the result shown as solid lines. The features of the curves such as the peaks and valleys become flattened with a filament that has a temperature gradient: The hotter central parts of the filament emit orders of magnitude less electrons than the colder edges of the filament, which obscures the emission of the center and flattens the valley of the curve. The y-axis indicating thermionic emission is in log scale.}
  \label{fig:TheoreticalAndSmeared}
\end{figure}

Taylor and Langmuir ~\cite{Taylor, Taylor2} describe a relationship between the temperature of a tungsten filament, the cesium vapor pressure, and tungsten thermionic emission. The dotted curves in Fig. \ref{fig:TheoreticalAndSmeared} show the relationship of the thermionic emission as the temperature of the filament changes for different cesium atom fluxes, and is reproduced from Fig. 15 in their 1933 paper ~\cite{Taylor}. 

\par
As an educational sidebar in a program to develop large-area alkali metal photocathodes, we have explored thermionic emission by building a gauge to measure the partial pressure of cesium in the pressure range between $10^{-5}$ and $10^{-3}$ Pa. Section~\ref{oper} describes the gauge construction, assembly and operation. The results are presented in the context of the Taylor and Langmuir model in Section~\ref{results}.  Section~\ref{lessonsLearned} describes lessons we have learned from this first generation gauge, and describes specific solutions to issues we encountered.  
\section{Gauge Assembly and Operation}
\label{oper}

\begin{figure}[H]
  \centering
  \includegraphics[height = 190pt]{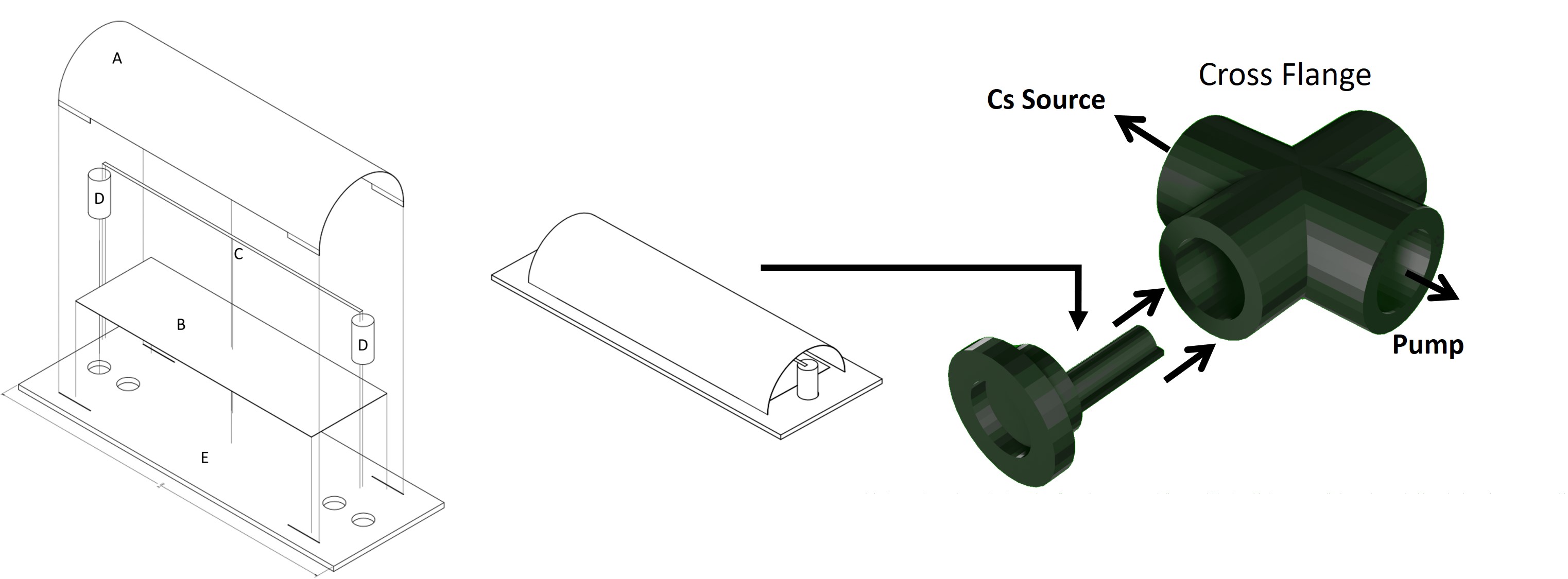}
  \caption{On the left, the exploded assembly of the gauge, on the middle, the assembled gauge, and on the right the positioning of the gauge in the manifold. There are copper wires that were not depicted connecting the cylindrical (A) and flat (B) copper plates, as well as the filament ends, to independent pins in a feedthrough. The filament (C) is connected to screws which act as leads (D). All elements are isolated from each other through ceramic beads and boards like the base ceramic support (E). The copper wires are thick enough to support the structure without it dipping and touching the manifold.}
  \label{fig:Detector}
\end{figure}

The proof-of-concept gauge is shown in Fig. \ref{fig:Detector}. It consists of a tungsten ribbon filament surrounded by a collector for the thermionic current. The collector comprises a flat base made with a ceramic rectangular plate coated with copper, on which is mounted a semi-cylindrical copper sheet. The tungsten ribbon is supported on the rectangular plate by isolating screws at each end which serve as terminals for the filament. All components are inside a 2 3/4" CF 4-way cross flange, and are individually connected by ceramic-isolated copper wires to a vacuum feedthrough at one end of the cross. The CF cross connects to a valve leading to a custom cesium source containing a glass vial of pure cesium~\cite{alfa_aesar}, and also connects to a turbo pump. After the system was pumped to a pressure in the order of 1.3x$10^{-5}$ Pa , the vial was broken to introduce the cesium to the system ~\cite{patent}. The pump was valved off during measurements. 

\par
Thermal control of the manifold is necessary to achieve temperature uniformity and control cold spots in the manifold where cesium vapor could condense. The manifold is surrounded by K-type thermocouples and heaters, which are covered by fiberglass insulation. The temperature of the cesium source was varied to check the measurements from the gauge against the calculated pressure.
\par

\begin{figure}
  \centering
  \includegraphics[height = 250pt]{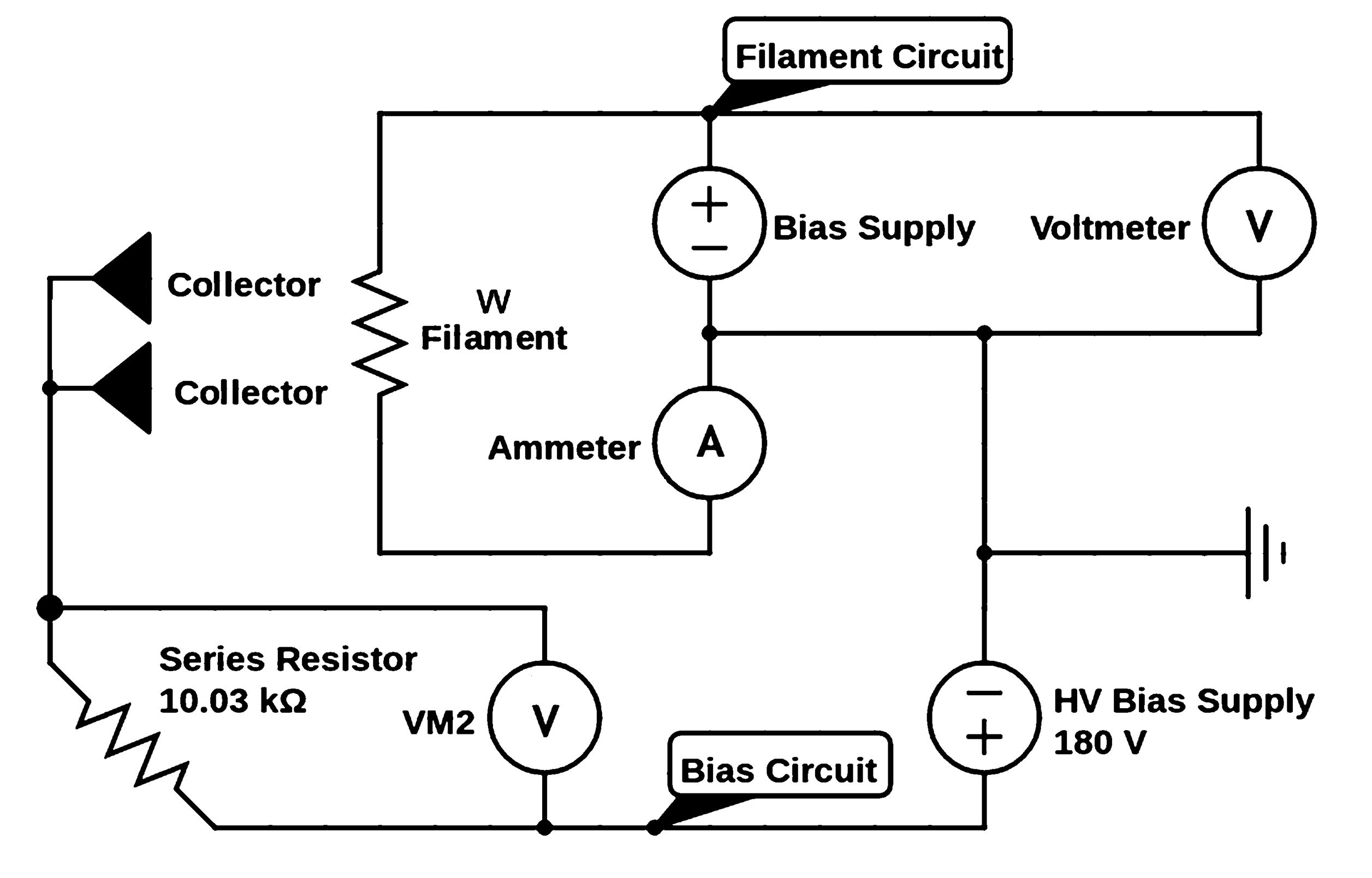}
  \caption{Circuitry of the gauge. Thermionic electrons leave the filament at ground voltage and are pulled by an electric field towards the collectors at a high voltage ($>$200 V+). They then pass through the series resistor and the voltage is  voltmeter 2 is proportional to the thermionic emission from the filament.}
  \label{fig:circuit}
\end{figure}

The circuit used to measure thermionic emission is outlined in Fig. \ref{fig:circuit}. It comprises two subsystems, the heating of the filament, and collectors that measure thermionic emission. The heating circuit consists of a power supply with built-in ammeter and voltmeter (A and V in Fig. \ref{fig:circuit}) connected to the screws at the ends of the tungsten filament, with the negative terminal set at manifold ground. The measurement circuit consists of the copper plate and semi-circular sheet collectors tied to a bias voltage relative to ground through a 10.03 k$\Omega$ resistor. The voltage from the collector current, typically 0.01 - 100 V, is measured across this resistor.

\par
\begin{figure}
  \centering
  \includegraphics[height = 300pt]{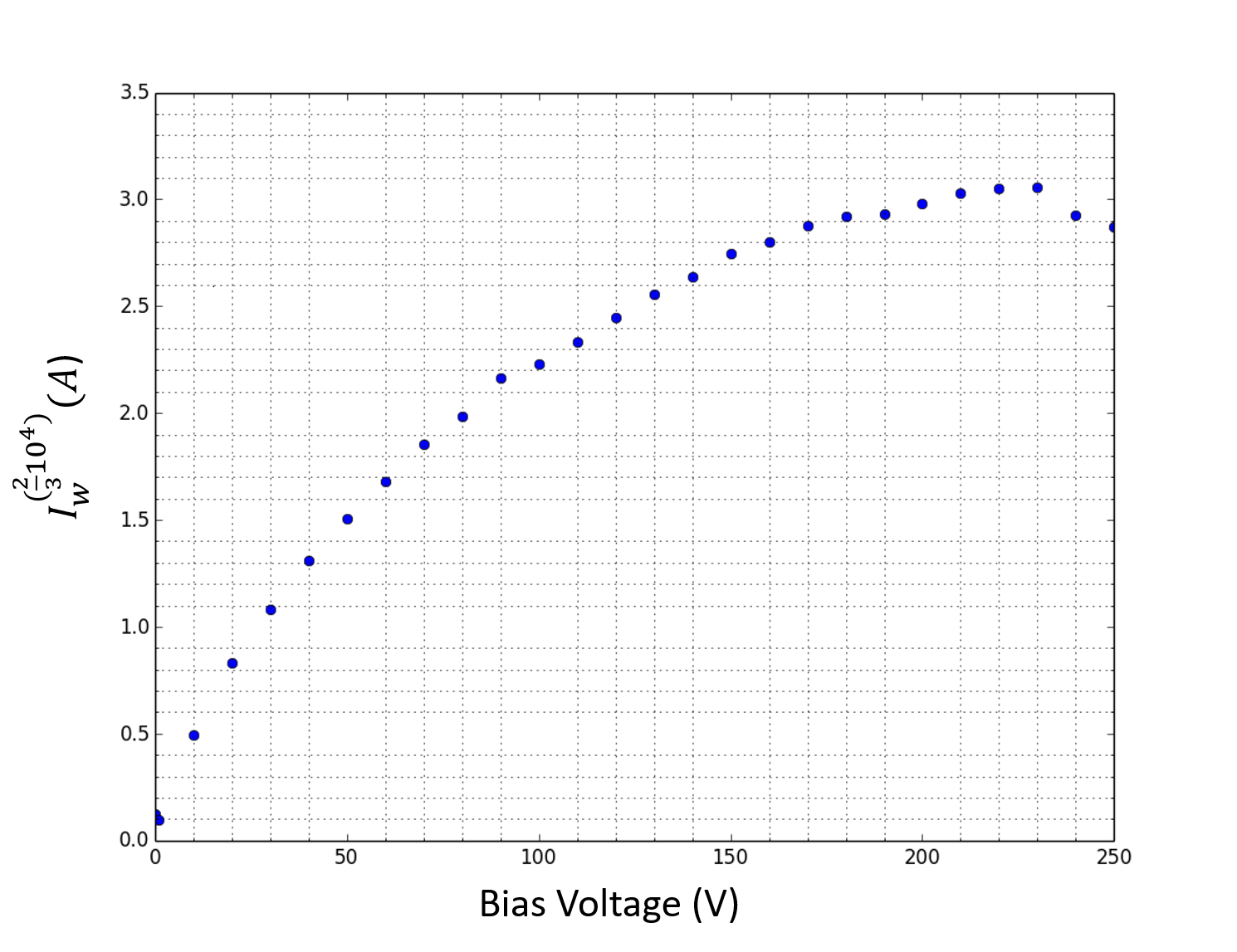}
  \caption{Curve of measured thermionic current versus increased bias voltage (difference between collectors and filament). In this case the critical voltage observed was around 200V and after that the collected current started plateauing. If measurements were taken at a voltage before the plateau, not all electrons would be collected for some temperatures of the filament which would indicate a lower pressure. The current was manipulated as the y-axis indicates to be in the same form as Fig. 21 in the Taylor and Langmuir paper ~\cite{Taylor}.}
  \label{fig:Plateu}
\end{figure}

 As the bias voltage increases from zero more thermionic electrons are directed to the collectors. There is a critical bias voltage after which the current plateaus because all emitted electrons reach the filament. We reproduced  Fig. 21 in the Taylor and Langmuir paper ~\cite{Taylor} to find this point, 180 - 200 V, as shown in Fig. \ref{fig:Plateu}. All measurements were performed at or beyond this critical bias voltage.
 
\par 

A measurement with the gauge consists of changing the the filament temperature by changing the current, and measuring the collection current by the voltage through the 10.03 k$\Omega$ resistor. The collected electron current divided by the surface area of the filament yields the thermionic electron flux. The conversion between filament current to filament temperature invokes a temperature transport equation, as described in Appendix A. 
\par
The measurements can be fit to the model of Taylor and Langmuir, which is used to infer the flux of cesium atoms in the manifold, and hence the Cs partial pressure. Measurements were taken without the aid of automated logging using voltmeters with 0.01 V precision. An automated measurement algorithm could easily be implemented.
\par 
Before each measurement the circuit was probed for shorts between circuit elements and/or ground from cesium condensation on the gauge surfaces. To eliminate the shorts, we applied 300 V between each element and between the elements and ground. The initial low resistances increased to over 10 k$\Omega$ once the bias was applied. If the system was left idle for more than two hours, even if heated, the shorts returned and the process had to be repeated.
\par

\section{Results and Discussion}
\label{results}
\begin{figure}[!ht]
  \centering
  \includegraphics[height = 400pt]{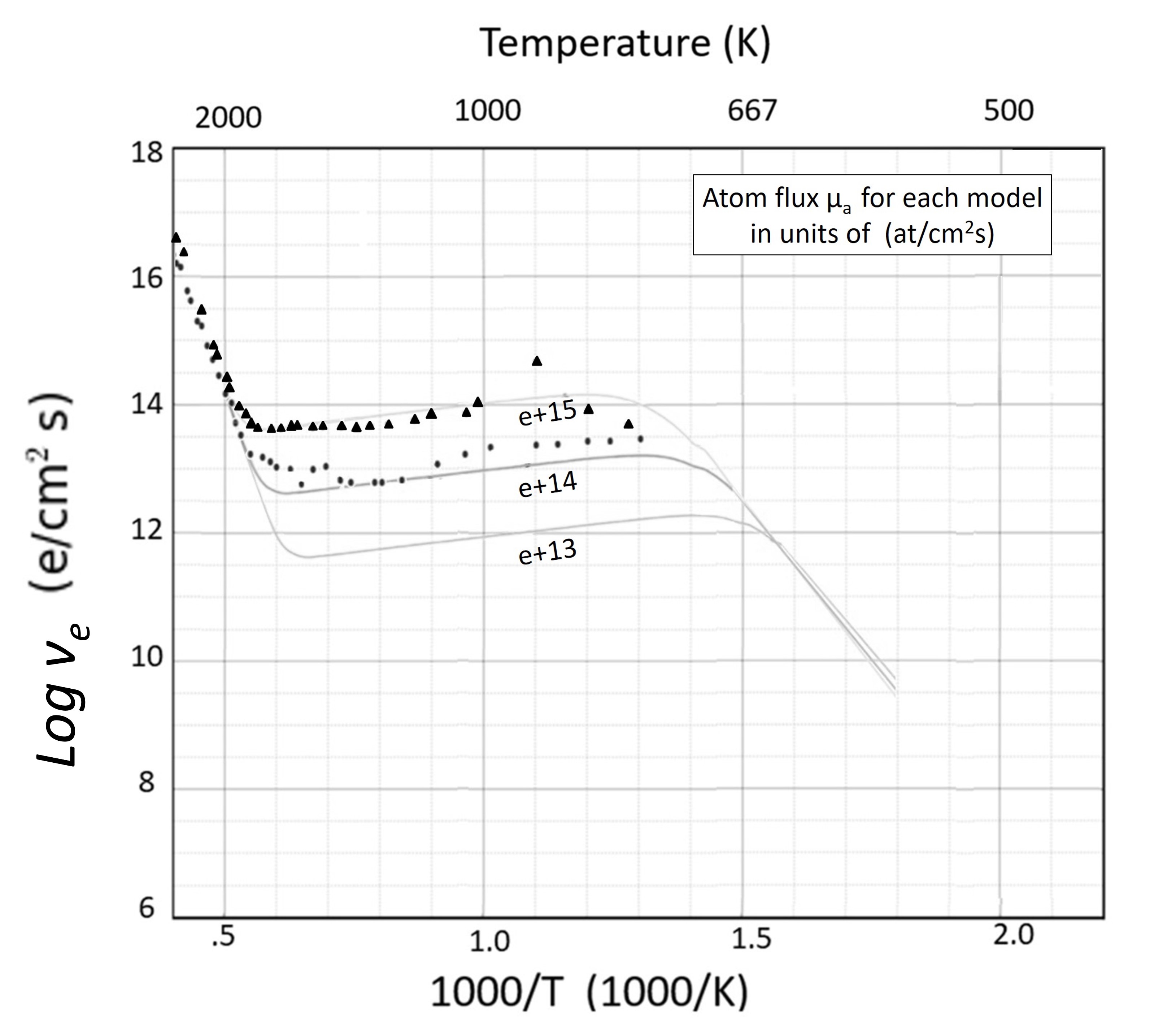}
  \caption{The logarithm of Thermionic emission as measured by the gauge versus 1000/Filament Temperature for two measurements. Measurement A (triangles) was taken at a higher partial pressure of cesium than measurement B (circles), with the cesium source temperatures being 363K and 305K respectively. The outlier in measurement A is most likely a mistake.}
  \label{fig:scurve}
\end{figure}

Measurements of thermionic emission were taken at source temperatures of 363 K (Measurement A) and 305 K (Measurement B) with the manifold temperatures at 503 K and 516 K, respectively. The measured pressures fit to $1.30$x$ 10^{-4}$ and $8.34 $x$ 10^{-4}$ Pa as shown in Fig. \ref{fig:scurve}. As expected, thermionic emission is greater at higher source temperatures. The results match a pure tungsten curve at high filament temperatures. 
\par

\begin{figure}[!ht]
  \centering
  \includegraphics[height = 400pt]{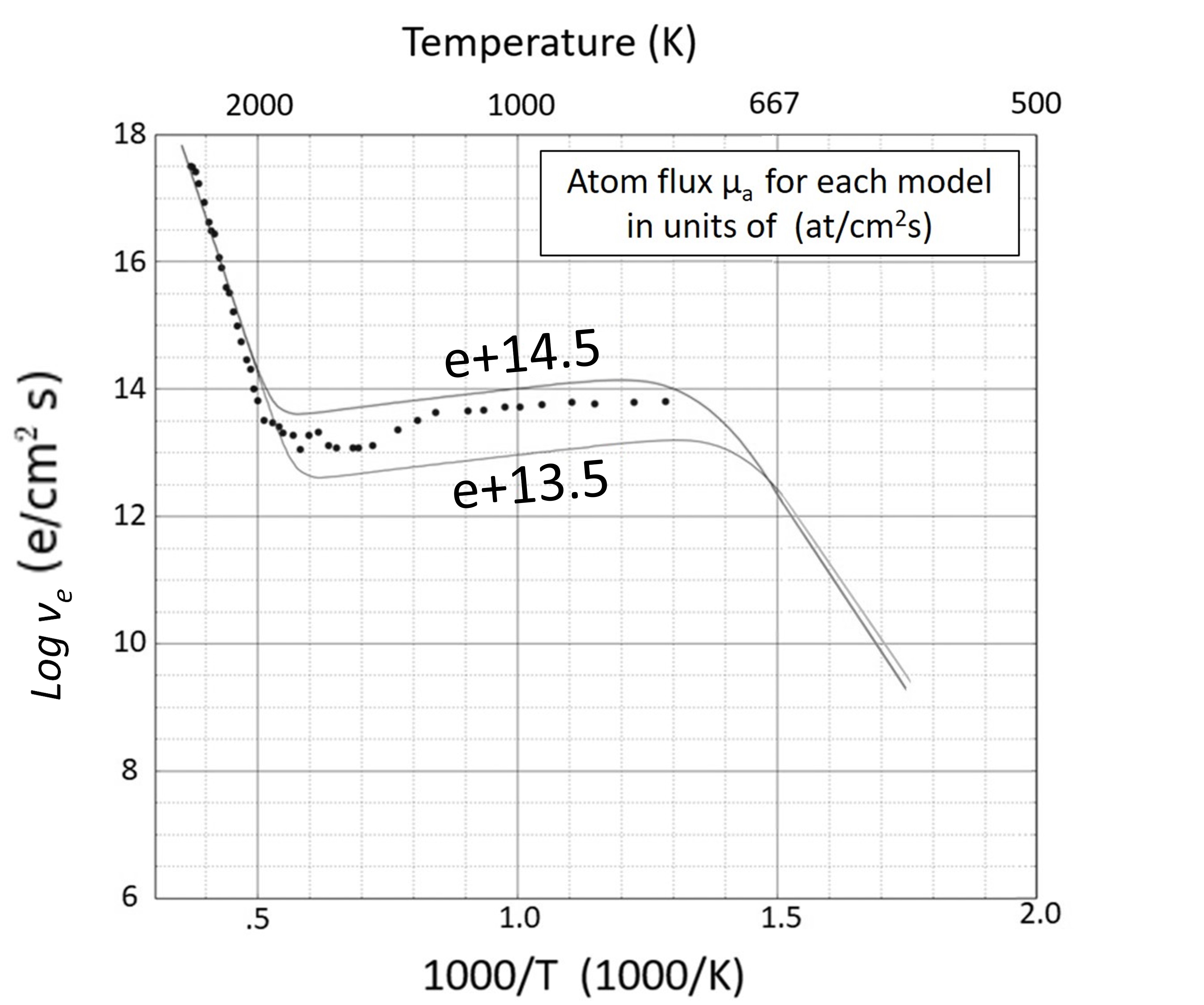}
  \caption{One measurement of the thermionic emission curve as a function of temperature. This measurement is fit to a solid curve corresponding to $10^{14.23}at/cm^2s$, an atom flux which can be converted to $1.30$x$ 10^{-4}$ Pa. The y-axis is in Log scale.}
  \label{fig:scurveTheo}
\end{figure}

The measured data do not match Taylor and Langmuir`s (TL) predictions. The filament we used had a temperature gradient due to thermal conduction at the leads. In contrast, the TL results were obtained by using guard rings to suppress collection from filament segments that were not uniform in temperature: electronic thermionic emission was nearly uniform along the fraction of the filament measured. A correction based on the temperature gradient was computed, described in Appendix A. The corrected curve can be seen in Fig. \ref{fig:TheoreticalAndSmeared} as the solid lines, and in Figs. \ref{fig:scurve},\ref{fig:scurveTheo}. The corrected model agrees much better with the data.

 \par 
At filament temperatures decrease towards that of the manifold, heat conduction into the filament becomes significant, altering the relationship between filament current and temperature. The gauge was operated at elevated temperatures ($>$200 $^{\circ}$C, above 500 K for the measurements) to minimize cesium condensation; results at filament temperatures close to that of the manifold are omitted.
 
 \par
The proof-of-principle gauge failed after commissioning and two series of measurements due to cesium condensation impervious to evaporation. After the system had been opened and exposed to air, we verified a darkening of essentially all ceramic surfaces. Between electronic terminals, lighter patches appeared. This coloration pattern can be explained by cesium deposition and later vaporization between electronic terminals. When exposed to air, remaining cesium darkened. For a more in-depth discussion on operation precautions and leakage currents, see Appendix B.

\section{Lessons Learned: Proposed Second Generation gauge} 
\label{lessonsLearned}
\begin{figure}[H]
  \centering
  \includegraphics[height = 300pt]{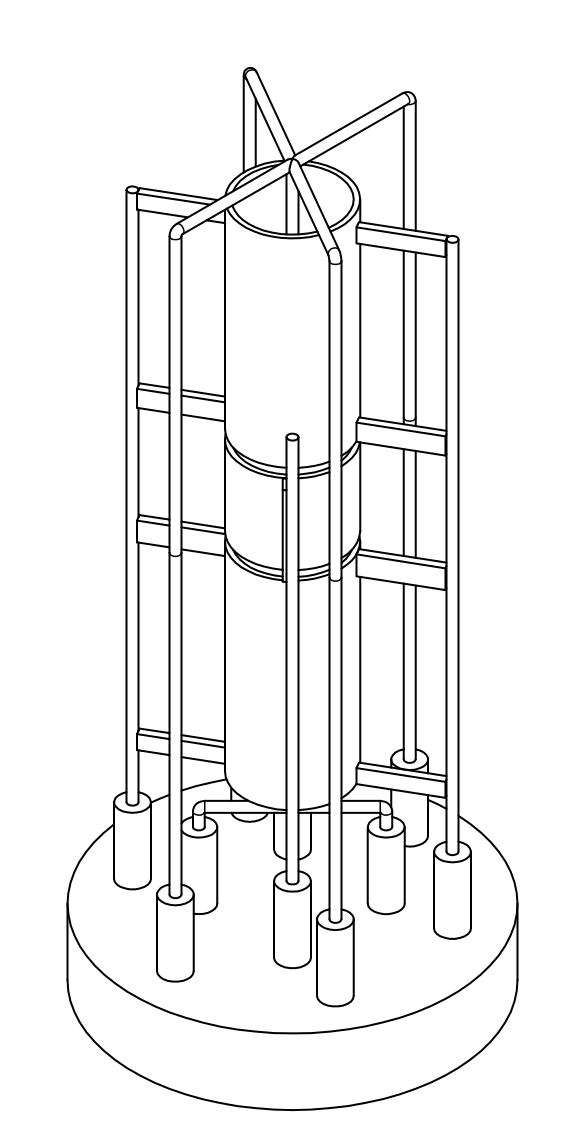}
  \caption{A scheme for a possible gauge redesign based on the results of the proof-of-concept gauge and its shortcomings. This design includes guard rings and minimizes structural-nonconducting surfaces to avoid leakage currents and improve signal.}
  \label{fig:new}
\end{figure}

The proof-of-concept gauge demonstrated the Taylor and Langmuir model can be used to measure cesium pressures, but made apparent that we had made mistakes in the design. The condensation of cesium vapor shorting circuit elements, and the presence of a filament temperature gradient are the main issues. An example second-generation instrument is illustrated in Fig. \ref{fig:new} and described in Subsection A. Alternatively, Springer and Cameron had proposed another method using a Bayer-Alpert gauge to measure partial pressures of cesium using its ions instead of electrons ~\cite{BOB, privateCharlie}, as described in Subsection B.

\subsection{Improved gauge using thermionic emission}
\par 
In order to counter leakage currents and electrical shorts, the gauge must minimize areas for cesium deposition around relevant conductors such as the collector or the leads. All circuit elements must avoid contact with support structures if possible. One solution is to have the circuit elements be self-supporting as in Fig. \ref{fig:new}.
\par
\begin{figure}[H]
  \centering
  \includegraphics[height = 100pt]{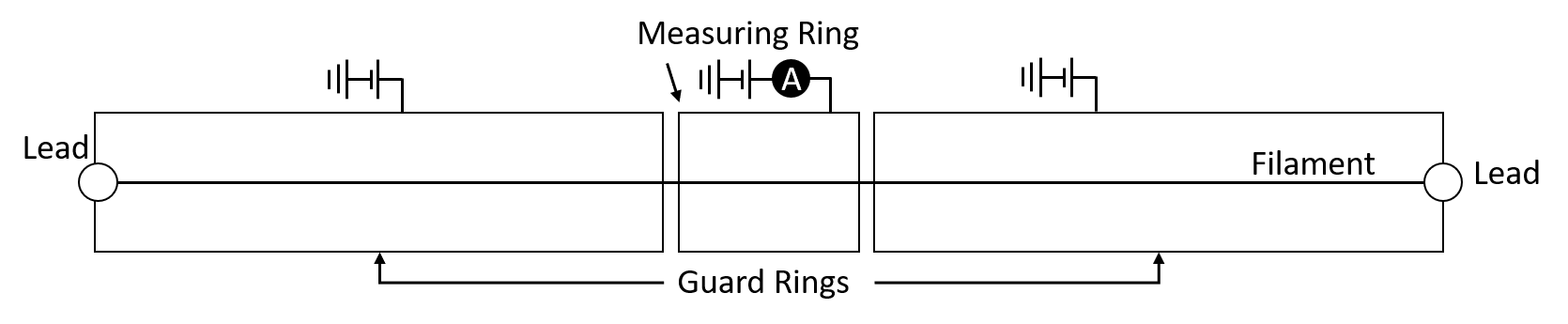}
  \caption{A scheme for how guard rings and the measurement conductor would be arranged. All three conductors would be biased at the same voltage and have the same shape. Around the center of the filament, the temperature can be considered almost constant. This means that instead of a flattened curve, we would measure the same curve Taylor and Langmuir predict. Cylinders attached to the base are made of ceramic/insulating material, but all other elements are metal. It is paramount that the setup is structurally rigid to avoid shorts from small mechanical impacts/deformations.}
  \label{fig:Guards}
\end{figure}

 
To account for the filament temperature gradient, the collector may be split into two guard rings and a central collector ring (Fig. \ref{fig:new}, Fig. \ref{fig:Guards}). All rings should be cylindrical and biased at the same voltage to create a uniform radial electric field. Only the central guard ring would be used to measure thermionic emission as the center is the hottest part of the filament with the smallest gradients. This was done by Taylor and Langmuir~\cite{Taylor}.

\par 
The effectiveness of the guard rings is improved if the ends of the filament are coiled, increasing the length of the filament and thus flattening the temperature gradient. Although the emissions from the coiled areas will not follow the Taylor and Langmuir model, they will be shielded by the guard rings and not measured by the central ring.

\subsection{Using ionized cesium over electrons as an alternative}
\par

An alternative method is to measure the current of ionized cesium ions instead of electrons thermionically emitted from the metal surface. A commercial Bayard Alpert gauge (BAG) is an ion gauge with an emitter filament, a collector filament a cylindrical grid along the collector length. Springer and Cameron heated the grid to eject electrons and ionize cesium and biased the collector filament negatively to collect the ions. Measuring the collected current yields the cesium partial pressure in a similar way to the method described in this paper. 
\par 
We had originally intended for the proof-of-principle gauge to be usable in ionic and electronic regimes. However, given the difficulties with cesium condensation the electron regime was preferred.

\section{Conclusion}
\label{conclusion}
From the relationship between the thermionic emission of a tungsten filament and its temperature in the presence of cesium vapor, a gauge can be built to determine the partial pressure of the cesium. We have shown a proof-of-concept gauge and measurements, documented its problems, and proposed a new design. We hope this paper is useful to those who wish to construct a similar device, or to explore the principles of thermionic emission, metallic crystal growth on metal surfaces, and work-functions. 

\section{Acknowledgements}
\label{acknowledgements}

The authors would like to thank (Dr.) Evan Angelico, Andrey Elagin, Henry Frisch and Mary Heintz for their help during the project; Matthew Poelker and Charles Sinclair for their insightful comments and helpful suggestions. We also thank Nicole Dombrowski, Hayward Melton and Hannah Tomlinson for their previous work on assembling the proof-of-concept gauge. 

\section*{Appendix A: Calculations for Wire Temperature Gradient}
\label{Mathematics}
To obtain the partial pressure of cesium, we must produce an experimental curve of thermionic emission versus temperature of the filament. To do so it is necessary to convert the two observable parameters, the current through the tungsten filament (\(I_w\)) and the current observed entering collecting conductors (\(I_c\)) to the temperature of the filament (\(T_w\)) and to the number of electrons leaving the filament (\(\nu_e\)) respectively.
\par
The temperature of the filament is paramount to the characterization of the thermionic current. In their experiments, Taylor and Langmuir made use of guard rings to only measure thermionic electrons from the very center of their filaments so that they may safely assume its temperature is uniform in that region. However, without using guard rings, and instead measuring electrons from the entire length of the filament, it is necessary to account for the filament's temperature gradient.
\par
Since the leads conduct heat away from the filament, there is a temperature gradient with a peak at the midpoint. The equation that represents the equilibrium (a steady temperature distribution) for a small part of the filament is:
\begin{equation}
    \centering
    0 = dPe + dPti - dPi - dPto
    \label{FirstTransport}
\end{equation}

\begin{figure}[H]
  \centering
  \includegraphics[height = 100pt]{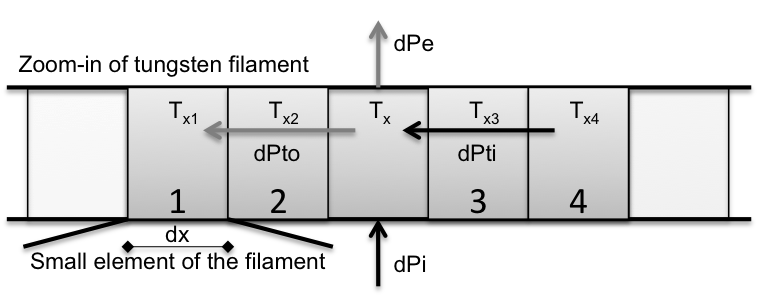}
  \caption{Power transport scheme for a small piece of the filament, the different pieces are numbered as they are used in the equations \ref{Transport2} to \ref{endEquation}. The Black arrows indicate power into the filament piece, and the grey arrows indicate power out. \(dPti\) is the power piece 3 conducts from piece 4 into the center piece, and \(dPto\) is the power piece 2 conducts from the center piece to piece 1.}
  \label{fig:FilScheme}
\end{figure}

Where \(dPe\) is the power lost by emission, \(dPi\) is the power from the current through the filament, \(dPti\) is the power conducted in from the hotter end and \(dPto\) is the power conducted out at the colder end (see Fig. \ref{fig:FilScheme}). Each of these components can be written out as a function of the position in the filament as follows:
\begin{equation}
    \centering
    dPe = dA_e \epsilon_{(T_{(x)})} \sigma T_{(x)}^4 
    \label{Transport2}
\end{equation}    
\begin{equation}
    \centering
    dPi = \frac{i^2 \rho_{(T_{(x)})} dx}{A_t}
\end{equation}
\begin{equation}
   \centering
    dPti = A_t K_{(T_{(x_3)})} \frac{T_{(x_4)} - T_{(x)}}{dx}
\end{equation}
\begin{equation}
    \centering
    dPto = A_t K_{(T_{(x_2)})} \frac{T_{(x)} - T_{(x_1)}}{dx}
\end{equation}
\par
Where \(l, w\) and \(d\) are the dimensions of the filament, with \(l\) being its long edge, \(A_e\) is the external area of the filament , \(A_t\) is the transverse area of the filament (in our case of a flat filament, \(A_e = 2l(w+d)\) and \(A_t = wd\) respectively); \(\sigma\) is the Stefan-Boltzmann constant, \(T_{(x)}\) is the temperature gradient; \(i\) is the current through the filament. Furthermore,  \(\rho_{(T_{(x)})}\) is the resistivity of tungsten as a function of temperature; \(\epsilon_{(T_{(x)})}\) is the emissivity of tungsten as a function of temperature; and \(K_{(T_{(x)})}\) is the conductivity of tungsten as a function of temperature. These three tungsten properties were all fitted to 4th or 5th order polynomials as a function of T.

\par
Note that equations 5 and 6 can be re-written as a function of the first derivative of \(T_{(x)}\) at \(x_3\) and \(x_2\) respectively:

\begin{equation}
    \centering
    dPti = A_t K_{(T_{(x_3)})} \frac{dT(x_3)}{dx}
\end{equation}
\begin{equation}
    \centering
    dPto = A_t K_{(T_{(x_2)})} \frac{dT(x_2)}{dx}
\end{equation}
 Substituting equations 3, 4, 7 and 8 back into equation 2 we have:
 \begin{equation}
    \centering
    0 = dA_e \epsilon_{(T_{(x)})} \sigma T_{(x)}^4 + \frac{i^2 \rho_{(T_{(x)})} dx}{A_t} + (A_t K_{(T_{(x_3)})} \frac{dT(x_3)}{dx} - A_t K_{(T_{(x_2)})} \frac{dT(x_2)}{dx})
\end{equation}
Which can also be re-written as a function of the second derivative of \(T_{(x)}\):
 \begin{equation}
    \centering
    0 = dA_e \epsilon_{(T_{(x)})} \sigma T_{(x)}^4 + \frac{i^2 \rho_{(T_{(x)})} dx}{A_t} + A_t ((\frac{dT_{(x)}}{dx})^2\frac{dK_{(T_{(x)})}}{dT_{(x)}} + K_{(T_{(x)})}\frac{dT_{(x)}^2}{d^2x})dx
    \label{endEquation}
\end{equation}    
\par
This is a second order differential equation that can be solved numerically for \(T_{x}\) with two boundary conditions. We assume knowledge of the temperature of the ends of the filament, which is valid if the leads it is connected to are heat-sunk enough to the manifold, the temperature of which we can measure. Furthermore, given that at high temperatures \(K_{(T)}\) is small, we assume that for the very center of the wire, where it is hottest, equation 2 has negligible conduction summands, and thus we can solve for the temperature at the midpoint given a current. Therefore, we have both \(T_{(0)}\) and \(T_{(\frac{l}{2})}\) as boundary conditions. Solving for \(T_{(x)}\) we get, for any current, a temperature gradient in the filament.
\par
Once we have \(T_{(x)}\), we can perform a numerical integration of how many thermionic electrons will be emitted by different parts of the filament based on Taylor and Langmuir`s experimentally determined curve. We do this to calculate the predicted thermionic emission from a filament with non uniform temperatures, and we can generate a curve analogous to Taylor and Langmuir`s but for a filament with a temperature gradient instead of a uniformly heated filament (See dotted lines in Fig. \ref{fig:TheoreticalAndSmeared}).

\section*{Appendix B: Operation Precautions}
A measurement with the gauge has been described above. However, to obtain reproducible results, certain precautions must be taken.
\par
Taylor and Langmuir recommend that before any measurement is done the filament undergoes what the call an aging process. This involves leaving the filament heated at 2400K for 10 hours, then at 2600 K for an hour and finally conclude with a few brief flashes at 2900K. Prior to such aging, neither we nor Taylor and Langmuir could obtain reproducible results. Furthermore, we have determined that brief flashes to about 2500K before any measurement generate more reproducible results as well.

\par 
It is also necessary that the measurement not interfere appreciably with the temperature of the source of cesium. The filament heats up to high temperatures during the measurement (up to 2500 K) and if the cesium source temperature changes mid-measurement, so will the pressure of cesium vapor. A fast measurement can solve this issue, as well as placing the source far away from the filament or using a thinner filament.
\subsection{Leakage Current}

Leakage currents between the charged plates and either ground or the filament terminals increase the background in the measurement of the currents collected from the plates. We attribute this leakage current to condensed cesium on the ceramic tile connecting the ends of the filament, the plates and the manifold together.
\par 
There are two ways leakage currents interfere with the measurements. It makes it so that the signal measured across the series resistor does not only come from collected thermionic electrons, but also from ground electrons that arrive from leakage conductive paths. Furthermore, as one changes the temperature of the conductive paths, cesium evaporates or condenses, thus changing the resistance of the leakage pathways. This means that there is not a simple static current background one might subtract from the measured signal.
\par 
It is possible to deal with such currents, however. By applying high voltages to the circuit elements any conductive cesium paths to ground will heat up and the cesium will evaporate off the surfaces. For the charged plates, half an hour at 300V took a 10 k\(\Omega\) short to up to tens of M\(\Omega\). Another effective counter to leakage currents was to keep the gauge much hotter than the source to avoid creating sites prone to cesium deposition. When source and manifold were left at the about the same temperature, all of the circuit elements were connected to each other and to ground by resistances under one k\(\Omega\). When the manifold was left at a four times the temperature of the source, however, resistances went up to hundreds of k\(\Omega\).
\par


\end{document}